\documentclass[aps,prx,showpacs,superscriptaddress,longbibliography,twocolumn,nofootinbib]{revtex4-2} 

\usepackage{subcaption}
\usepackage[T1]{fontenc}
\usepackage[utf8]{inputenc}
\usepackage{amsmath,amsthm,amssymb,amsfonts}
\usepackage{bm,bbm}           
\usepackage{physics}      
\usepackage{amssymb}

\usepackage{graphicx}
\usepackage{float}
\usepackage{booktabs}     
\usepackage{multirow}

\usepackage{xcolor}
\definecolor{darkBlue}{rgb}{0,0,0.8}
\usepackage[colorlinks=true, allcolors=darkBlue]{hyperref}

\usepackage{microtype}     
\usepackage[ruled]{algorithm2e}
\usepackage[capitalize,nameinlink]{cleveref} 
\usepackage{lipsum}

\definecolor{darkGreen}{RGB}{0,110,0}

\begin{document}

\title{The dimensionality of the Hopfield model}

\author{Cristopher Erazo}
\email{cerazova@sissa.it}
\affiliation{Scuola Internazionale Superiore di Studi Avanzati (SISSA), Trieste, Italy}

\author{Santiago Acevedo}
\email{sacevedo@sissa.it}
\affiliation{Scuola Internazionale Superiore di Studi Avanzati (SISSA), Trieste, Italy}

\author{Alessandro Ingrosso}
\email{alessandro.ingrosso@donders.ru.nl}
\affiliation{Donders Centre for Neuroscience, Radboud University, Nijmegen, The Netherlands}
\date{\today}

\begin{abstract}
We use the Binary Intrinsic Dimension (BID), a geometrical measure designed for binary data, to analyze the Hopfield model, a paradigmatic spin system from statistical mechanics, machine learning and neuroscience. The BID allows us to characterize the phases and
transitions of this system, and moreover it is robust against finite-size effects that interfere with the correct numerical estimation of the spin-glass order parameter ($q$).
We observe that the BID scales linearly with system size in the retrieval and paramagnetic phases, where the correlations between spins are small, and exhibits sublinear scaling in the whole spin-glass phase, highlighting its correlated structure. Furthermore, we establish a direct relationship between the BID and the overlap distribution, unveiling a novel connection between the geometry of the state-space and standard spin order parameters.

\end{abstract}

\maketitle

\section{Introduction}

Complex systems exhibit emergent phenomena and phase transitions, often without clear or easily identifiable order parameters. 
Nonetheless, the specific interactions between the constituents of a physical system give rise to correlation patterns between features that shape a geometry in data space, such that the observed data live in a manifold $\mathcal{M}$~\cite{goldt-manifold,bengio-rep,isomap}. 
In recent years, the numerical estimation of the dimensionality of $\mathcal{M}$ has been extensively studied~\cite{CAMASTRA201626}, with fruitful applications in machine learning and data science~\cite{Laio_2019,cheng2025emergence,ID-diffusion,ID-neuro}. The development of robust dimensionality estimators is crucial to gain insights from large-scale recordings of neural dynamics~\cite{Gao_2015,Cunningham_2014}, where low-dimensional structures appear to be conserved across brain regions in different species~\cite{perich2025neural}.

The traditional technique for dimensionality estimation in unsupervised learning is Principal Component Analysis (PCA), which consists in analyzing the empirical covariance matrix of the data and identifying a set of linear features that explain most of the variance. 
While PCA has been used to estimate dimensionality in an extremely wide range of scientific applications, it is inherently limited to linear manifolds and it overestimates the intrinsic dimension (ID) of curved, nonlinear data structures.
To address these limitations, local methods have been developed to estimate dimensionality based on the scaling of the volume around a point's neighborhood \cite{levina-bickel,Laio_2017}. These methods are better suited for curved manifolds and have been successfully applied to study the geometrical properties of data representations in neural networks, even predicting classification accuracy in various applications \cite{Laio_2019}. Local methods, however, suffer from the so-called \emph{curse of dimensionality}, as estimating local observables requires a number of samples that grows exponentially with the input dimension. This limitation leads to poor scaling and a severe underestimation of the ID in large dimensions, even for uncorrelated data~\cite{Acevedo_2025}.
To overcome these challenges,~\cite{Acevedo_2025} introduced the \emph{Binary Intrinsic Dimension} (BID), a dimensionality measure specifically designed for high-dimensional spin systems. Unlike local methods, the BID is a global observable that avoids the curse of dimensionality and provides an estimation of the intrinsic dimension that scales linearly with the number of spins for systems that can be effectively decomposed in non-interacting subsystems. 
On traditional two-dimensional statistical mechanics systems in thermal equilibrium, the BID has shown to be able to properly characterize global correlation structures of different phases of matter, identifying the corresponding phase transitions.
Moving away from thermal equilibrium, the BID was found to display universal exponents in one-dimensional critical dynamical systems modeling surface growth~\cite{Verdel_2025}.
In the context of machine learning, the BID has been used as a proxy for the semantic content of representations generated by convolutional networks trained for image classification, and to characterize the semantic correlations of text representations in Large Language Models~\cite{Acevedo_2025}.

In this work, we apply the BID to study the Hopfield model~\cite{amari1972learning,Hopfield_original}, the simplest example of an attractor network~\cite{Nishimori_2001} with binary units. The Hopfield model is a paradigmatic spin system that has been studied in the context of spin glasses and has recently gained renewed attention due to its connections with Restricted Boltzmann Machines~\cite{BARRA20121}, Transformer architectures~\cite{Ramsauer2020HopfieldNI}, and its role in modeling learning and generalization in machine learning~\cite{pham2024memorization,correlated_patterns_negri,Matteo_2}.
In contrast to traditional analytical studies focusing on the thermodynamic limit of the system~\cite{amit1987statistical,Amit_2}, we focus on the numerical study of a finite-size model, with sizes of the order of thousands of units, comparable to the embedding dimensions of modern deep learning architectures. 

\bigskip

The structure of our work is as follows: 
In~\cref{sec:BID} we introduce the BID, and how to measure it in generic spin systems. 
In~\cref{sec:Hopfield} we present the Hopfield model, its phases and order parameters. 
In~\cref{sec:results} we show our results, first measuring the dimensionality of the model across its parameter space in~\cref{sec:bid_parameter_space}, showing that it can be used as an order parameter. In~\cref{sec:BID-and-q} we make some connections to the theory of spin glasses. In~\cref{sec:system_size} and~\cref{sec:scaling_exp} we focus on the finite-size scaling of the BID. Finally,~\cref{sec:conclusions} discusses conclusions and perspectives.

\section{Methodology}\label{sec:methd}
\subsection{Binary Intrinsic Dimension}
\label{sec:BID}
\begin{figure}[t]
    \centering
    \begin{subfigure}[t]{0.34\linewidth}
        \centering
        \caption{Dataset}
        $\mathcal{S} \in \{-1, +1\}^{N_S \times N}$
        \includegraphics[width=0.85\linewidth]{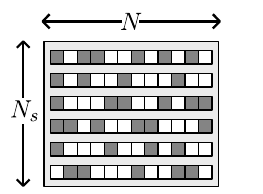}
        \label{fig:BID-ilustration(c)}
    \end{subfigure}
    \hfill
    \begin{subfigure}[t]{0.62\linewidth}
        \centering
        \caption{Hamming Distances}
        \includegraphics[width=1\linewidth]{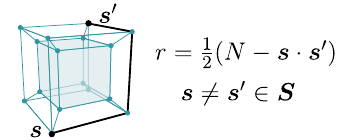}
        \label{fig:BID-ilustration(b)}      
    \end{subfigure}

\begin{subfigure}[t]{\linewidth}
        \caption{Binary Intrinsic Dimension}
        \centering
        \begin{minipage}[t]{0.49\linewidth}
            \vspace{0pt} 
            \centering
            \makebox[\textwidth][l]{%
            \includegraphics[width=1.0\linewidth]{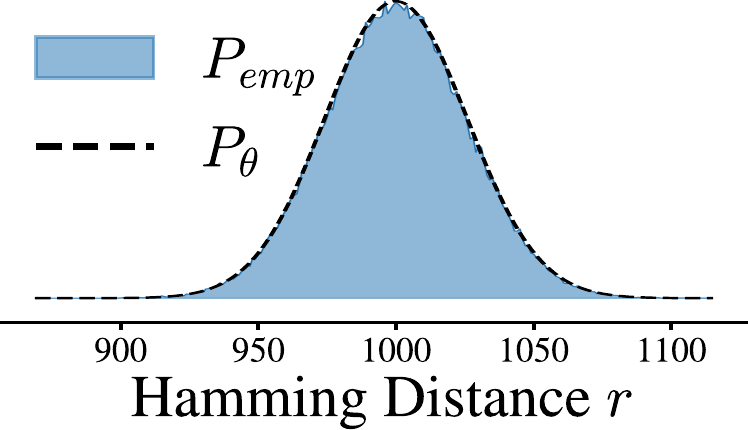}
            } 
        \end{minipage}
        \hfill
        \hfill
        \begin{minipage}[t]{0.45\linewidth}
            \vspace{0pt} 
            \centering
            Ansatz: 
            \vspace{4pt}
            $d_{\theta}(r) = \theta_0 + r\theta_1$\\
            $\downarrow$\\
            \vspace{4pt}
            $\hat{\theta} = \arg\displaystyle\min_{\theta} D_{KL}(P_{emp}||P_\theta)$\\
            $\Rightarrow\;\;\; BID = \hat{\theta}_0$
        \end{minipage}
        \label{fig:BID-ilustration(c)}
    \end{subfigure}

    \caption{\raggedright (a) Binary Dataset. (b) Illustration of Hamming distance between $\bm s$ and $\bm s'$ as the length of the shortest path connecting both spins in the $N$ dimensional hypercube. (c) The BID is obtained by fitting the model \cref{eq:bid_model} to the empirical distribution of Hamming distances \(\{r(\bm s,\bm s'): \forall \bm s\neq\bm s' \in \mathcal{S}\} \).}
    \label{fig:BID-ilustration}
\end{figure}

The Binary Intrinsic Dimension (BID) is a measure of the intrinsic dimensionality of a dataset, specifically designed for high-dimensional systems with binary degrees of freedom. 
To define it, let us first consider a system of \( N \) non-interacting binary spins or neurons. Spin configurations \(\bm{s} \in \{-1,+1\}^N\) are then uniformly distributed across the vertices of an \( N \)-dimensional hypercube (\cref{fig:BID-ilustration(b)}). The Hamming distance between two configurations, defined as \( r = \frac{1}{2}(N - \bm{s} \cdot \bm{s'}) \), has the exact distribution
\begin{equation}
    P_0(r) = \dfrac{1}{2^N} {N \choose r}.
    \label{eq:uncorr_spins}
\end{equation}
For uncorrelated spins, the intrinsic dimension is exactly \( N \). However, in real datasets, correlations among the degrees of freedom reduce the effective dimensionality of the system, and the empirical distribution of Hamming distances deviates from \( P_0(r) \). To account for these correlations, \cite{Acevedo_2025} proposed the following model for the distribution: 
\begin{equation}
    P(r|\theta) = \dfrac{1}{Z}\dfrac{1}{2^{d_\theta(r)}}{d_\theta(r)\choose r}
    \label{eq:bid_model}
\end{equation} 
where \( Z \) is a normalization constant, and \( d_\theta(r) \) is an analytic, scale-dependent dimensionality measure. The normalization is computed as \(Z=\sum_{r\in \mathcal{R}} \frac{1}{2^{d_\theta(r)}}{d_\theta(r)\choose r}\) where the range \(\mathcal{R} \subset \{0,\cdots,N\}\) contains the set of all sampled distances.

The model reduces to the uncorrelated case when \( d_\theta(r) = N \) and it has been observed in practice that the distribution \( P(r|\theta) \) can accurately fit the empirical distribution \( P_{emp}(r) \) by performing a first-order Taylor expansion for \( d_\theta(r) \), such that

\begin{equation}
    d_\theta(r) = \theta_0 + r\theta_1. 
    \label{eq:d(r)_linear}
\end{equation}
The optimal parameters \( \hat{\theta} = (\hat{\theta}_0, \hat{\theta}_1) \) are obtained by minimizing the Kullback-Leibler divergence between the empirical distribution and the model. For practical reasons, in most cases we perform a local fit of the model only on a subset of \(\mathcal{R}\), typically focusing on distances ranging from the smallest measured distances up to a cutoff \(r_{max}\) where the linear model works best (see \cref{subsec:gradient,subsec:range_R} for details). The Binary Intrinsic Dimension is then defined as the limit of small distances of \(d_\theta(r)\): 
\[ BID = \hat{\theta}_0. \]

For convenience, we work with the \emph{normalized overlap variable} \( x = \frac{1}{N} \bm{s} \cdot \bm{s'} \), which is linearly related to the Hamming distance: 
\begin{equation}
 x = 1-\frac{2}{N}r \quad \leftrightarrow \quad r=\frac{N}{2}(1-x).
\label{eq:transform_r_x}
\end{equation}
This transformation allows us to express the BID and other quantities in terms of \( x \), 
a central quantity in spin glass theory. In this work, we use $\mathbb{E}_{emp}[x]$ and $\mathbb{E}_\theta[x]$ to denote the expected value of $x$ over the empirical distribution and the model, respectively, and likewise for the standard deviation.\\

\subsection{Hopfield Network}
\label{sec:Hopfield}

The Hopfield model is a system of \( N \) binary neurons or spins \(\bm{s} \in \{-1,+1\}^N\) coupled pairwise by the Hamiltonian: 
\[ H(\bm{s}) = -\dfrac{1}{2} \sum_{i,j=1}^N J_{ij} s_i s_j. \] 
The couplings \( J_{ij} \) are constructed to store \( p \) binary patterns \(\bm{\xi}^\nu \in \{-1,+1\}^N\) (\( \nu = 1, \dots, p \) ) using the Hebbian rule: 
\[ J_{ij} = \dfrac{1}{N} \sum_{\nu=1}^p \xi_i^\nu \xi_j^\nu, \quad J_{ii} = 0. \] 

These couplings ensure that the patterns are local minima of the Hamiltonian and that the system will act as an associative memory by converging to a pattern $\bm{\xi}^\nu$ when initialized close to it. The \emph{load} of the network is defined as \(\alpha=p/N\). The dynamics of the network follows asynchronous updates, where each spin \( s_i \) is updated sequentially according to the conditional probability: 

\[ P(s_i = +1 \mid \bm{s}_{\setminus i}) = \frac{1}{1 + \exp(-2\beta h_i)}, \] 
where \( \beta = 1/T \) is the inverse temperature, and the local field is computed as: 
\[ h_i = \sum_{j \neq i} J_{ij} s_j. \] 
This dynamics is equivalent to Gibbs sampling from the Boltzmann distribution: 
\begin{equation}
    \mu(\bm s) = \dfrac{e^{-\beta H(\bm s)}}{Z_\beta}.
    \label{eq:boltzmann-measure}
\end{equation}

In the thermodynamic limit (\( N, p \to \infty \) with \( \alpha \) fixed), the Hopfield model exhibits three distinct phases, characterized by the spin-glass order parameter \( q \) and the overlaps (magnetizations) \( m^\nu \) with the stored patterns: 
\begin{equation}
    q = \frac{1}{N}\displaystyle\sum_{i=1}^N\langle s_i\rangle^2  \quad \text{and}\quad
m^\nu = \frac{1}{N}\displaystyle\sum_{i=1}^N\langle s_i\rangle\xi^\nu_i 
\label{eq:ord-params}
\end{equation} 
Here, \( \langle \cdot \rangle \) denotes the thermal average over \( \mu(\bm{s}) \).\\

The phases of the model in the thermodynamic limit are the following.

\begin{itemize}
    \item \textbf{Retrieval (Stable)}: the patterns are the \emph{global} minima of the free energy, and the system spontaneously aligns with one of the patterns $ (m^\mu > 0, q = 0, m^\nu = 0$ for $\nu \neq \mu$ ).
    
    \item \textbf{Retrieval (Metastable)}: the patterns are only \emph{local} minima of the free energy, with a glassy state being the global minima. The system will thus align with one of the patterns only when it starts sufficiently close to it. The transition line separating the metastable retrieval and stable retrieval phases is called $T_c(\alpha)$, and corresponds to a first order phase transition.
   
    \item \textbf{Spin glass}: for high memory load and low temperature, retrieval states are no longer local minima of the free energy, the only global minimum being a glassy state. Individual neurons will develop disordered spontaneous polarizations, but not in the direction of the patterns. This phase is characterized by $ q > 0$ and $ m^\mu = 0$, $\forall \mu $. The transition line between the metastable retrieval and spin glass phases, $T_M(\alpha)$ is also a first-order phase transition.
    
    \item \textbf{Paramagnetic}: for high enough temperatures there is a single minimum of the free energy, and neurons are uncorrelated (\( m = 0, q = 0 \)). The transition line between the paramagnetic and spin glass phases, $T_g(\alpha)$, is a second-order phase transition. 
\end{itemize}

\begin{figure*}
    \begin{subfigure}[ht]{0.49\linewidth} 
    \caption{\raggedright}
        \includegraphics[width=1.0\linewidth]{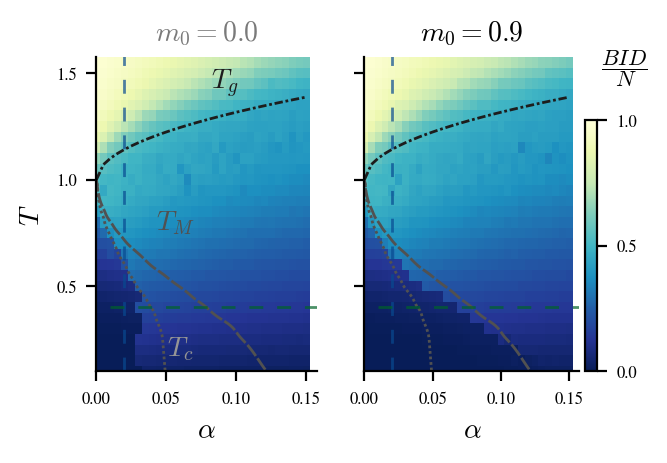}
    \label{fig:BID-ord-param(a)}
    \end{subfigure}
 \hfill
        \begin{subfigure}[ht]{0.49\linewidth}

    \caption{ \raggedright}
        \includegraphics[width=1.0\linewidth]{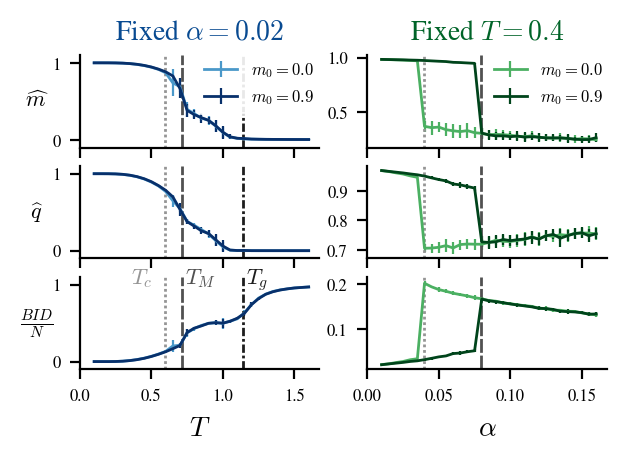}
        \label{fig:BID-ord-param(b)}
    \end{subfigure}
    \caption{\raggedright (a) Mapping of the parameter space using the BID for two initial \emph{magnetization} values \(m_0 = 0.0\) and \(m_0=0.9\). The transition temperatures are indicated by their corresponding labels. Vertical and horizontal dashed lines represent cuts at fixed $\alpha$ and $T$, respectively. (b) Numerical Hopfield order parameters \( (\widehat{q},\widehat{m})\) computed at fixed value of $\alpha$ and $T$ alongside the $BID$, normalized per bit. The vertical dashed lines correspond to the transition points identified in panel (a), with different line-styles indicating the transitions.  Simulations were performed with $N=1024$, $N_S=2500$, averages are computed over $30$ realizations.}
    \label{fig:BID-ord-param}
\end{figure*}

The transition temperatures between phases, called \(T_g\), \(T_M\) and \(T_C\) are displayed in the parameter space of \cref{fig:BID-ord-param(a)} and were found solving numerically the fixed point equations for the minimization of the free energy of the system in the thermodynamic limit, obtained by the replica method~\cite{Nishimori_2001}.

For a finite $N$ system, the thermal average \(\langle\cdot \rangle\) is estimated from a finite dataset \(\bm{S} = \{\bm{s}^{(t)}\}_{t=1}^{N_S}\) of samples generated by the dynamics. For example, the empirical estimates of \( q \) and \(m^\nu\) are given by: 

\begin{equation}
\begin{split}
    \widehat{q} &= \frac{1}{N} \sum_{i=1}^N \left( \frac{1}{N_S} \sum_{t=1}^{N_S} s_i^{(t)} \right)^2 \\
    \widehat{m}^\nu &= \frac{1}{N} \sum_{i=1}^N \left( \frac{1}{N_S} \sum_{t=1}^{N_S} s_i^{(t)} \right) \xi_i^\nu
\end{split}
\label{eq:order-params-estimates}
\end{equation}

It is important to distinguish between these numerical estimates computed at finite size and the theoretical values of \( q \) and \( m^\nu \), computed in the limit of infinite system size (\( N \to \infty \)). In the thermodynamic limit, the system is self-averaging, and the disorder average coincides with the thermal average of \cref{eq:ord-params}: 

\begin{equation} 
q = \lim_{N \to \infty} \mathbb{E}_{\bm{\xi}} \left[ \frac{1}{N} \sum_{i=1}^N \langle s_i \rangle_{\bm \xi}^2 \right],
\end{equation}
where $\langle s_i \rangle_{\bm \xi}^2$ stands for the squared thermal average of $s_i$, with a fixed realization of patterns $\boldsymbol{\xi}$.
This distinction will be crucial in the results section, where we analyze the finite-size effects and demonstrate how the BID provides a robust estimation across system sizes. In \cref{subsec:averages} we provide further details on the nomenclature related to averages.

\section{Results}
\label{sec:results}

\subsection{$BID$ over parameter space}
\label{sec:bid_parameter_space}

The first key result of this work is the demonstration that the BID behaves as an effective order parameter for the Hopfield model, capturing the transitions between the different phases. In \cref{fig:BID-ord-param}, we compute the BID across the parameter space, characterized by the temperature \(T\) and the load $\alpha$.
\cref{fig:BID-ord-param(a)} shows the BID across the parameter space for two different \emph{initial magnetization} values, \(m_0 = 0\) and \(m_0 = 0.9\). The value of \(m_0\) determines the fraction of initial spins that are set equal to those of one pattern, therefore these initial conditions correspond to random initialization and initialization close to the selected pattern, respectively. This choices allow us to distinguish between the stable and metastable regions of the retrieval phase.\footnote{If we set \(m_0=0\), the typical overlap between the initial random random configuration and the pattern is still \(\bm s^{(0)}\cdot\bm\xi^0 \approx\frac{1}{\sqrt{N}}\).}

For \(m_0 = 0\), the BID is close to zero in the retrieval phase (\(T < T_C\)), consistent with the fact that the dynamics converges to a single stored pattern, resulting in a dataset of nearly identical configurations, up to small thermal fluctuations. For \(m_0 = 0.9\), the small-dimensionality extends up to \(T \approx T_M\), reflecting the metastable nature of this region. In this case, the system can converge to a stored pattern only if the initial configuration is within its basin of attraction. In the spin-glass phase (\(T_M < T < T_g\)), the BID takes intermediate values, reflecting the disordered but correlated nature of the system. Finally, in the paramagnetic phase (\(T > T_g\)), the BID increases to \(\frac{BID}{N} \approx 1\), indicating that the spins are uncorrelated.

In \cref{fig:BID-ord-param(b)}, we project the BID along vertical and horizontal cuts in the parameter space (fixed \(\alpha\) and \(T\), respectively). Alongside the BID, we plot the traditional Hopfield order parameters \(\hat q\) and \(\hat m\), estimated numerically using \cref{eq:order-params-estimates}. The vertical dashed lines correspond to the theoretical transition points identified in \cref{fig:BID-ord-param(a)}. 
 
Note that the phase diagram depicted in \cref{fig:BID-ord-param} is computed numerically with a system of \(N=1024\) spins, which is a relatively small value to contrast against the thermodynamic solution. The selection of this system size is crucial to be able to observe the stable retrieval region for \(m_0=0\), as the waiting time to attain a magnetization with one pattern from purely random initialization becomes computationally prohibitive for large systems.
On the other hand, such a small size system is subject to stronger finite size effects. The most evident effect is observed in Fig.~\ref{fig:BID-ord-param}, where the transition lines computed for the infinite system are slightly shifted with respect to the
largest changes in the BID.
A more subtle effect is observed in the plot \(\hat m\) vs \(T\) of \cref{fig:BID-ord-param(b)} (blue curves), where we expect a vanishing magnetization in all the spin glass region, but instead we observe a smooth decrease to zero. 
Furthermore, in the \(\hat q\) vs \(T\) plot the spin glass order parameter drops to zero for \(T\approx1.0\) instead of \(T_g = 1.12\). This early vanishing of \(\hat q\) is observed consistently for several values of \(\alpha\) (see \cref{fig:mapping-ord-par(a)}) and is an important finite size effect that will be discussed in depth in the next sections. In the following we will always use values of \(\alpha = 0.04\) and \(m_0=0.9\) for the experiments.

\subsection{BID and distribution of overlaps \(x\)}
\label{sec:BID-and-q}

As discussed in \cref{sec:methd}, the BID is a global observable computed from the distribution of Hamming distances \( r = \frac{1}{2}(N - \bm{s} \cdot \bm{s'}) \) between pairs of spin configurations. This can also be expressed in terms of the normalized overlap variable \( x \), as defined in \cref{eq:transform_r_x}, which allows us to explore the relationship of the BID with the spin-glass order parameter \(q\).

\begin{figure}
    \centering
    \includegraphics[width=1\linewidth]{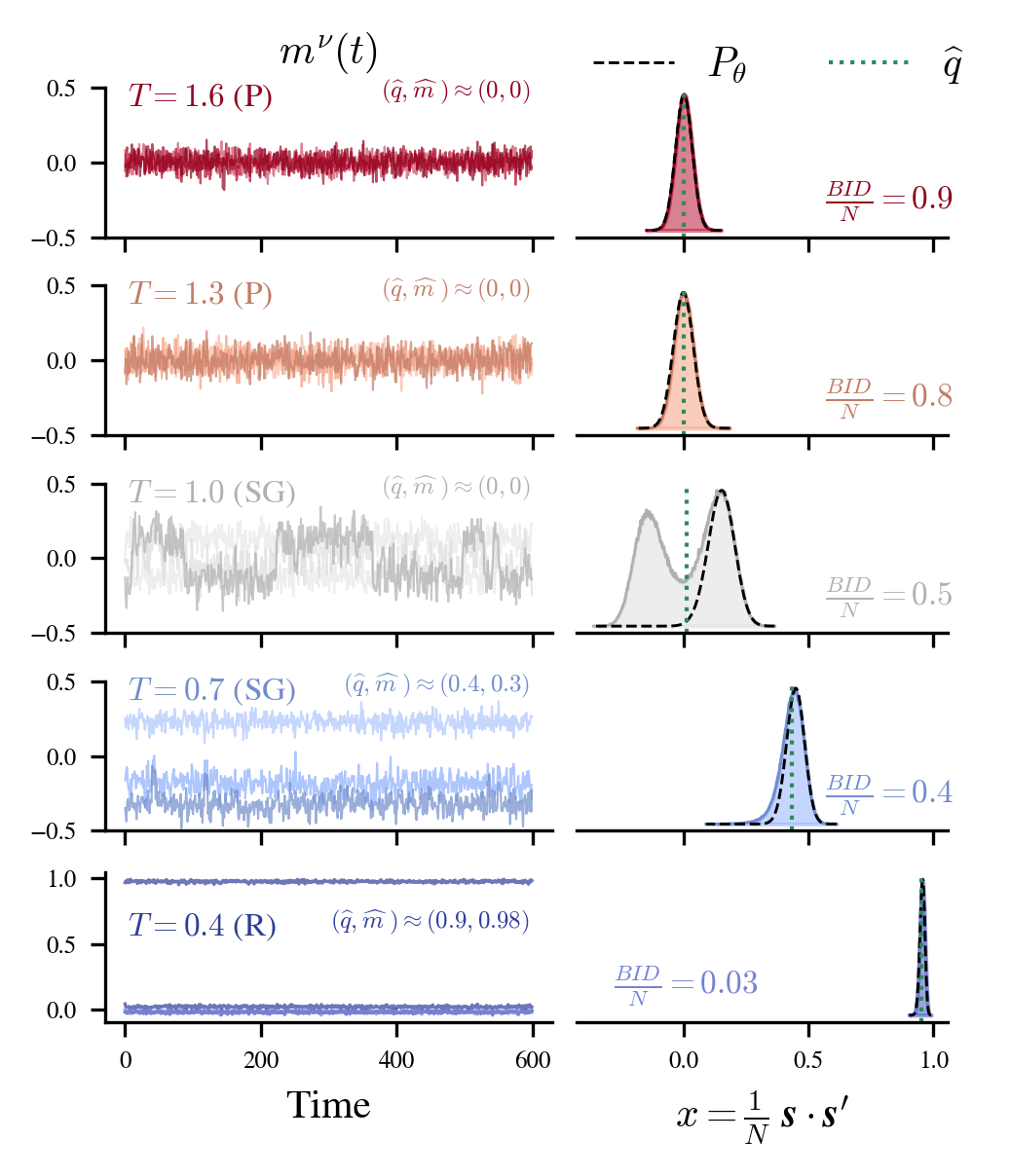}
    \caption{\raggedright Magnetization with the different patterns during asynchronous dynamics. For each temperature, the traces of several \(m^\nu(t)\) are shown along with the corresponding histogram of distances. In the corner of each trace plot we show the numerical order parameters \((\hat q,\hat m)\) estimated by~\cref{eq:order-params-estimates}. The value of \(\hat{q}\) is also shown in the histograms as the dotted vertical line. Simulation with \(N=1024\) and \(\alpha=0.04\).
    Note that in the spin glass phase, the model fit in black dashed lines is accurate for the right hand side of the probability distribution containing larger overlap values (intra-cluster distances). See~\cref{sec:BID-and-q} and~\cref{subsec:range_R} for further details. }
    \label{fig:mag-hist}
\end{figure}

In \cref{fig:mag-hist}, we show the overlap \( \hat   m^\nu(t) \) of the spin state with the stored patterns during asynchronous dynamics, alongside the empirical histograms of the normalized overlap \( x \) at different temperatures, for a fixed value of $\alpha = 0.04$. The fitted model distribution \cref{eq:bid_model} is shown as the dashed black line. These plots yield interesting insights into the behavior of the system across different phases:

\textbf{Retrieval Phase} (\( T = 0.4 \)): The magnetization \( \hat m^\nu(t) \) shows that the system strongly aligns with one of the stored patterns (\( \hat m \approx 1 \)), while the overlaps with other patterns remain close to zero. This is reflected in the histogram of \( x \), which is sharply peaked near \( x = 1 \), indicating that all configurations are nearly identical. The BID captures this low-dimensional structure effectively.

\textbf{Paramagnetic Phase} (\( T = 1.3 \) and \( T = 1.6 \)): The magnetization traces are nearly random, with no significant alignment to any stored pattern (\( \hat m \approx 0 \)). The histogram of \( x \) is centered around \( x = 0 \), consistent with uncorrelated configurations. At high temperatures, the distribution will approach the uncorrelated case \(P_0\) described by \cref{eq:uncorr_spins}.

\textbf{Spin-Glass Phase} (\( T = 0.7 \) and \( T = 1.0 \)): At \( T = 0.7 \), this finite size dynamics reaches a spurious state where the values of \(\hat m^\nu\) are non zero, but there is no pattern with perfect magnetization as in the retrieval phase. The histogram of \( x \) is biased and centered in the range \( x \in (0, 1) \), reflecting the correlated but disordered nature of the spin-glass phase. At \( T = 1.0 \), the dynamics exhibit sudden jumps between positive and negative magnetization values, a manifestation of the \( \mathbb{Z}_2 \) symmetry of the Hamiltonian (\( H(-\bm{s}) = H(\bm{s}) \)). The symmetry implies that the dataset can be decomposed in two clusters of states \(\mathcal{S} = \mathcal{S}^+ \cup \mathcal{S}^-\) resulting in a bimodal distribution of \( x \). There is one peak corresponding to small Hamming distances (intra-cluster distances within \( \mathcal{S}^+ \) or \( \mathcal{S}^- \)) and the other to large Hamming distances (inter-cluster distances between \( \mathcal{S}^+ \) and \( \mathcal{S}^- \)).

The vertical dotted lines in the histograms of \cref{fig:mag-hist} indicate the numerical value of \( \hat q \), computed using \cref{eq:order-params-estimates} from the dataset \(\mathcal{S}\). Its precise value is also displayed at the end of the trace plots along with \(\hat m\). By definition, the value of \(\hat{q} \) is the mean of the empirical distribution of \( x \) when \(N_S\) is large enough (see \cref{subsec:averages} for the derivations). 
At \( T = 1.0 \), the value of \(\hat{ q} \) is already close to zero, even though the theoretical transition temperature for \(\alpha = 0.04\) is \( T_g = 1.2 \). This discrepancy arises due to the \(\mathbb{Z}_2\) symmetry of the system, which causes the empirical average to vanish for small system sizes. We present the finite-size scaling of the histogram of distances in the next section. 

\begin{figure*}[ht]
    \begin{subfigure}[t]{0.48\linewidth} 
    \caption{\raggedright}
     \includegraphics[width=1.0\linewidth]{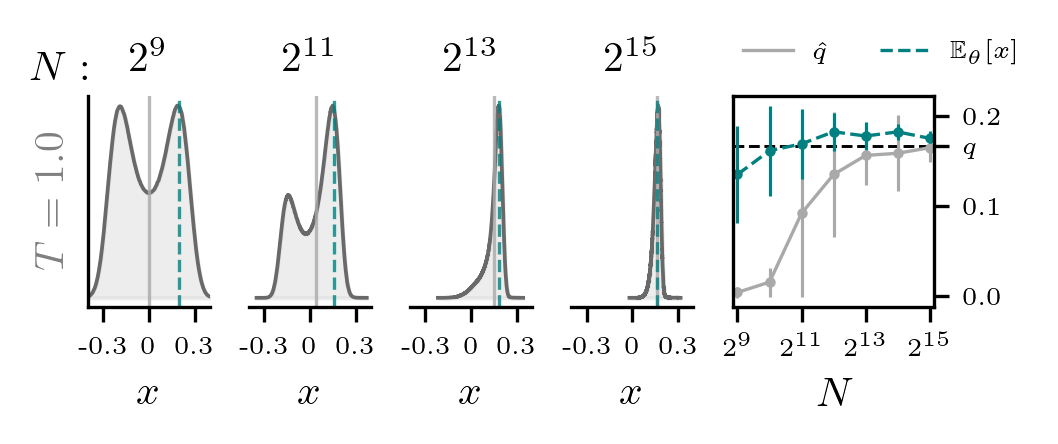}     
    \label{fig:scaling-N(a)}
    \end{subfigure}
    \hfill
    \begin{subfigure}[t]{0.48\linewidth} 
        \caption{\raggedright}
        \label{fig:scaling-N(b)}
        \includegraphics[width=1.0\linewidth]{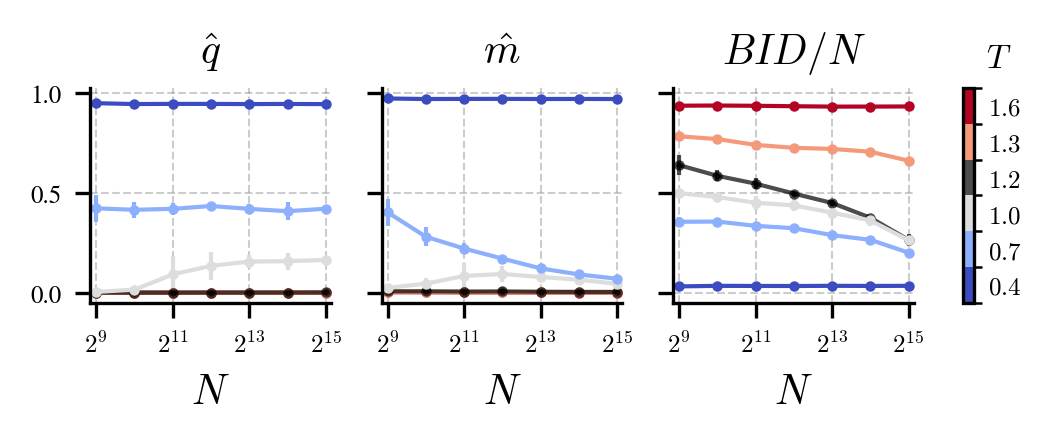}      
    \end{subfigure}   
     \caption{\raggedright (a) Typical histograms at temperature $T=1.0$ for different system sizes showing the bimodality as a finite size effect. 
     The last panel shows in gray the estimated $\hat q$ as a function of \(N\) and in teal the expected value of $x$ computed with the model \cref{eq:bid_model} after fitting.  Error bars correspond to the standard deviation across 20 independent network realizations.
     The thermodynamic value \(q \approx 0.17\) is shown as the horizontal dashed line.  In the histograms the vertical lines correspond to the values of \(\hat q\) and \(\mathbb{E}_\theta[x]\) in gray and teal color respectively. (b) Scaling of the order parameters with $N$ for different temperatures. The temperatures are the same as in figure \cref{fig:mag-hist} except for $T=1.2$ which is the critical temperature $T_g$ for $\alpha=0.04$. The curve of \(\hat q\) for \(T=1.0\) is the same one displayed on the last panel of (a). All simulations are done with \(\alpha=0.04\) and $m_0=0.9$.}
    \label{fig:scaling-N}
\end{figure*}

The BID, however, can overcome the symmetry issue. The model in \cref{eq:bid_model} is uni-modal and therefore, to fit the distribution we focus on the peak corresponding to small Hamming distances (large \(x\)). The selection of the range allows us to choose the region containing mainly intra-cluster distances, estimating correctly the dimensionality of each dynamical component.

This allows the BID to correctly capture the geometry within the clusters, and allows the model $P_\theta$ to properly fit only one mode of the empirical distribution of distances, overcoming the finite-size effects observed in the estimation of the mean, \( \hat{q} \).
As we will see later in \cref{sec:system_size}, the average overlap under \cref{eq:bid_model}, \(\mathbb{E}_\theta[x]\), is a better proxy for the thermodynamic value \(q\) than the Monte-Carlo estimate \(\hat{q}\), and in \cref{sec:scaling_exp} we use a Gaussian approximation of the model in \cref{eq:bid_model} to obtain an explicit relation between the BID and \(q\).

\subsection{Effects of system size}
\label{sec:system_size}

An example of the strongest finite size effect observed in the system is presented in \cref{fig:scaling-N(a)}, where we show the empirical distribution of \(x\) for fixed \(T=1.0\) and \(\alpha=0.04\), computed with increasing values of \(N\). We note that for small system sizes, the distribution is bimodal: the network alternates between two symmetrical regions of the state space. As \(N\) increases, the transition time between these regions grows, the distribution becomes unimodal and it concentrates around the thermodynamic value \(q\approx0.17\).

The BID obtained from \cref{eq:bid_model} is, nonetheless, less sensitive to such finite-size effects, since we perform a local fit focusing on the peak of distances smaller than the cutoff $r_{max}$, adaptively chosen using the quantiles of the distribution of distances (see~\cref{subsec:range_R} for details). This is shown in the teal dashed line in \cref{fig:scaling-N(a)}, which displays the expected overlap \( \mathbb{E}_\theta \left[ x\right]\) predicted by the model in \cref{eq:bid_model}. Unlike the empirical \(\hat q\), which is strongly affected by the bimodality, the model \cref{eq:bid_model} provides a consistent estimation across system sizes that matches very well the theoretical value at the thermodynamic limit.

In \cref{fig:scaling-N(b)}, we analyze the scaling of the BID and the Hopfield order parameters as a function of the system size \(N\) for different temperatures. For \(\alpha = 0.04\), the order parameters \(\hat{q}\) and \(\hat m\) are zero for \(T \geq 1.2\) (the critical temperature \(T_g\)) across all system sizes. In the retrieval phase (\(T = 0.4\)), both \(\hat q\) and \(\hat m\) are close to one, indicating strong alignment with one of the stored patterns. However, finite-size effects are evident in the spin-glass phase (\(T = 0.7\) and \(T = 1.0\)). For \(T = 1.0\), \(\hat q\) approaches zero for small system sizes (\(N \lesssim 1024\)) but increases with \(N\), stabilizing to a finite value. Similarly, for \(T = 0.7\) and \(T=1.0\), the magnetization decreases slowly to zero for large system sizes (\(N \gtrsim 4096\)).

The BID, as an intrinsic dimensionality measure, is expected to scale linearly with \(N\) for independent degrees of freedom. For two independent systems \(A\) and \(B\), the BID of the combined system should satisfy: 

\begin{equation}
    BID([A, B]) = BID(A) + BID(B),
    \label{eq:linear_ID}
\end{equation}
where \([A, B]\) represents the composite system. 
However, if \(A\) and \(B\) interact, the scaling is slower than linear due to the correlations induced by the interactions, which reduce the effective dimensionality of the combined system. 
This heuristic principle is reflected in the distinct scaling regimes observed in \cref{fig:scaling-N(b)}:

\textbf{Paramagnetic Phase} (\(T > T_g=1.2\)):
At high enough temperatures, \(BID/N \approx 1\) and remains approximately constant with respect to \(N\). 
This indicates that the system is weakly correlated, with degrees of freedom that are asymptotically independent as \(T\) increases.
Despite the all-to-all interactions in the Hopfield model, the high temperature suppresses macroscopic correlations, making the system effectively extensive.

\textbf{Retrieval Phase} (\(T < T_M\approx 0.6\)):
At low temperatures, \(BID/N \approx 0\) and is also constant with respect to \(N\). The small dimensionality reflects the system's magnetization with a stored memory, where the dynamics is dominated by its basin of attraction. The spins are trivially correlated due to the global alignment with the pattern, but the fluctuations around the pattern are effectively independent, resulting in a low-dimensional system.

\begin{figure*}[t]
    \centering
    \begin{subfigure}[t]{0.49\linewidth} 
    \caption{\raggedright} \label{fig:scaling_exponents(a)}
    \includegraphics[width=1.0\linewidth]{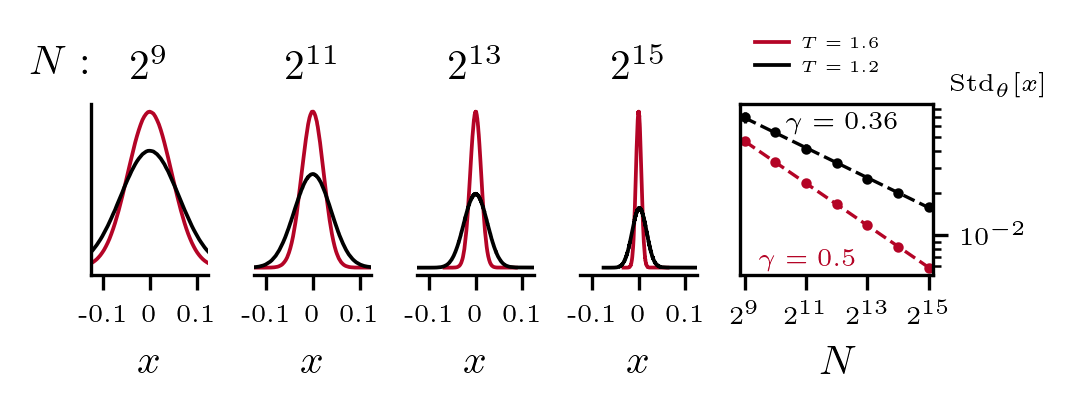}
    \end{subfigure}
    \hfill
     \begin{subfigure}[t]{0.49\linewidth} 
     \caption{\raggedright} \label{fig:scaling_exponents(b)}
         \includegraphics[width=0.7\linewidth]{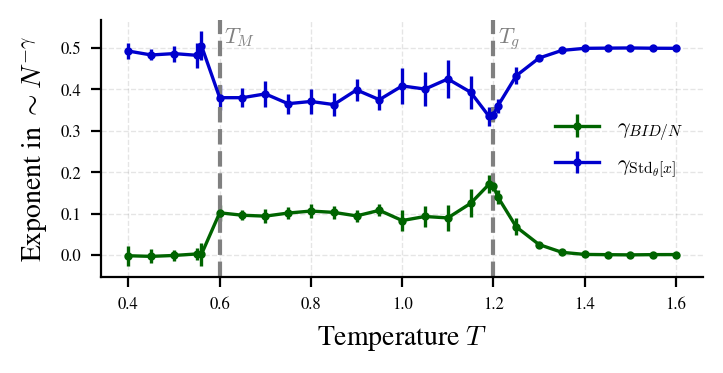}        
     \end{subfigure}
    
    \caption{\raggedright (a) Typical histograms at two temperatures for various $N$. The last panel shows the standard deviation of the distribution as a function of $N$ for each temperature along with the exponent \(\gamma \) in \(\text{Std}_\theta[x]\sim N^{-\gamma}\). (b) Scaling exponents for the BID and  \( \text{Std}_\theta\left[x\right]\) as a function of temperature. To obtain the value of \(\gamma\) and its dispersion we performed linear regression with errors on the values \(\frac{\text{BID}}{N}\), \(\text{Std}_{\theta}[x]\) vs. \(N\) for \(2^{10} \leq N \leq 2^{14}\). Experiments with \(\alpha=0.04\).}
    \label{fig:scaling_exponents}
\end{figure*}

\textbf{Spin-Glass Phase} (\(T_M < T < T_g\)):
In this phase, the BID exhibits sublinear scaling. This is more pronounced close to $T_g$ (black line), similarly to what was observed around the critical temperature of the ferromagnetic 2D-Ising model ~\cite{Acevedo_2025}. 
This scaling is a consequence of the collective correlations involving the whole system that break \cref{eq:linear_ID}. Nonetheless, away from $T_g$ and deep into the spin glass phase (\(T=1.0\) and \(T=0.7\) on \cref{fig:scaling-N(b)}), the scaling is still sublinear, which suggests that the correlations in this phase are restricting the dynamics to an ever decreasing fraction of the state space, effectively reducing the dimensionality of the explored landscape as \(N\) increases.

\subsection{Size scaling exponents}
\label{sec:scaling_exp}
In this section we show that the sublinear behavior of the BID is directly related to the scaling of the standard deviation of the overlap distribution. In \cref{fig:scaling_exponents(a)}, we plot the empirical histograms as a function of \(x\) for temperatures in two phases, and increasing values of \(N\). We can qualitatively observe that the concentration of the distribution towards the mean is quicker for the larger temperature. In the last panel of \cref{fig:scaling_exponents(a)} we display \( \text{Std}_\theta\left[x\right]\) as a function of \(N\) for both temperatures in log-log plot: we observe that the scaling exponents vary significantly from one case to the other.

In the paramagnetic phase,  \( \text{Std}_\theta\left[x\right] \sim N^{-1/2}\), consistent with the standard deviation of the sum of \(N\) independent random variables, and signaling the approximate independence between the degrees of freedom. However, in the spin-glass phase, \( \text{Std}_\theta\left[x\right]\) decreases with \(\gamma<1/2\), indicating that the central limit theorem does not apply anymore and the correlations among spins change the observed scaling laws, yielding a slower concentration towards the mean. To understand how this phenomena relates to the different scaling of the BID, we derived the following Gaussian approximation of \( P(r|\theta) \) in the large \(N\) limit: 
\begin{equation}
    P(r|\theta) \propto \exp\left(-\dfrac{(r-r_0)^2}{2\sigma_r^2} \right),
    \label{eq:P_gauss}
\end{equation}
where
\begin{equation}
     r_0 = \dfrac{\theta_0}{2 - \theta_1}, \quad \sigma_r^2 = \dfrac{2\theta_0}{(2 - \theta_1)^3}.
     \label{eq:gauss_par_theta}
\end{equation}

The approximation works very well for a large region of the parameter space (see \cref{subsec:Gaussian_apx}) and is useful because it provides a direct relationship between the BID and the summary statistics of the distribution. If we isolate \(\text{BID} = \theta_0\) from \cref{eq:gauss_par_theta} and transform the statistics using \cref{eq:transform_r_x} we obtain:

\begin{equation}
    \dfrac{BID}{N} = \dfrac{1}{\sqrt{N}\; \text{Std}_\theta\left[x\right]}\Big(1- \underbrace{ \mathbb{E}_\theta\left[x\right]}_{\approx \;q}\Big)^{3/2}.
    \label{eq:bid-xstats}
\end{equation}

From this equation, it is clear that the extensive behavior of the BID is associated with  \( \text{Std}_\theta\left[x\right] \sim N^{-1/2}\), while lower exponent for \(\text{Std}_\theta\left[x\right]\) leads to sublinear scaling of the dimensionality. 

To understand the scaling behavior of the system in the different phases, in \cref{fig:scaling_exponents(b)} we show the exponent \(\gamma\) for both the BID and standard deviation as a function of the temperature. The difference between the phases is also evident by looking at the corresponding exponents. For large temperatures, the BID scales linearly and \(\gamma_{\text{Std}} \rightarrow1/2\). This is expected since the Boltzmann measure \(\mu(\bm s)\) becomes approximately uniform over the spin configurations and the distribution of distances approaches \cref{eq:uncorr_spins}, for which the standard deviation is exactly \(\text{Std}_\theta \left[x\right]  =  N^{-1/2} \). 

When the temperature is decreased, correlations build up and the exponents change smoothly up to the critical value \(T_g\), where \(\gamma_{\text{Std}}\) is minimal and \(\gamma_{BID/N}\) maximal. On the other extreme, for the retrieval phase, we have similar scalings to the ones at \(T\rightarrow\infty\): spins behave almost independently, driven by the effective field aligned to the retrieved pattern \(\bm \xi\).

Finally, in the spin glass region, both exponents have an approximate plateau at an intermediate value that signals the non-trivial correlations among the degrees of freedom. Note that the error bars on the exponents are larger for \(1.0 \leq T \leq 1.2\), which is precisely the range of temperatures where we observed the bimodality of the distribution. This could affect the optimization algorithm, especially for small \(N\).

The general behavior of the two curves on \cref{fig:scaling_exponents} also confirms the validity of \cref{eq:bid-xstats}, which predicts \(\gamma_{\text{Std}} + \gamma_{BID/N} \approx 1/2 \). Using this equation, we can understand that the scaling behavior of the BID is determined by the variance of the distribution, which in turn reflects the correlation structure of the spins. On the other hand, the specific constant depends on the average overlap variable, which measures the alignment between the different configurations. 

We additionally checked that the BID sub-linearity observed in the spin glass phase is not simply a consequence of the time auto-correlation present in the Markov chain when collecting the samples. To do so, we simulated many independent chains with random initial conditions and constructed the dataset only with the last sample of each chain (see details in \cref{subsec:scaling_laws}). In this setup, the samples cannot have such autocorrelation time by construction, and we again find $\gamma_{BID/N} \approx 0.1 $ on the spin glass phase, showing that the sub-linearity is a consequence of the glassines of the system induced by the disordered couplings and the low temperature.

\section{Conclusions}
\label{sec:conclusions}

In this work, we demonstrated the utility of the Binary Intrinsic Dimension (BID) as a robust, unsupervised tool for characterizing the phases of the Hopfield model. Below, we summarize the main contributions, limitations, and future directions of this study.

\textbf{Main Contributions}

\emph{Unsupervised Detection of Phase Transitions:} We showed that the BID, a purely geometrical measure, can effectively identify the phase transitions in the Hopfield model without requiring prior knowledge of the system's dynamics or order parameters ($m$ and $q$). 

\emph{Relationship between BID and overlap \(q\):} 
In \cref{sec:BID-and-q} we established a direct relationship between the BID and the spin-glass overlap, showing that they are both encoded in the distribution of overlaps of the system. Our Gaussian approximation of the BID model in \cref{eq:bid-xstats} makes the connection between them explicit for the case of uni-modal distributions. 
Furthermore, we observe that the BID overcomes the finite-size effects present in the numerical estimation of the overlap \((\hat q)\) related to the $\mathbb{Z}_2$ symmetry of the Hamiltonian, that makes the distribution of overlaps bimodal.

\emph{Size-scaling of the BID:} We demonstrated that the BID scales linearly with the number of neurons in the retrieval phase and in the paramagnetic phase, reflecting their extensive nature due to trivial or vanishing correlations. In contrast, the BID exhibits sublinear scaling at the critical temperature $T_g$ and in the whole spin-glass phase, capturing the correlated structure of these two regimes. Such sub-linearity is not a consequence of the autocorrelation of a single Markov-Chain, but of the glassy nature of the system.

\textbf{Future Directions}

Future work involves exploring the application of the BID to more complex networks, such as dense associative networks and models with structured connectivity. These systems may exhibit richer phase diagrams and more intricate dynamics. 

Spiking networks, whose activity is binary in nature~\cite{pfeiffer2018deep}, are an interesting avenue for future work: both random, sparsely connected spiking networks~\cite{brunel2000dynamics}, as well as those trained using surrogate gradient methods~\cite{neftci2019surrogate,wang2020supervised}, could benefit from the BID's ability to characterize phase transitions and dynamical regimes. Similarly, the BID would be ideal to study binary neural networks
~\cite{hubara2018quantized,courbariaux2016binarized}, which are designed to reduce computational costs, saving time, memory and energy.

An interesting next step is to extend the BID framework to continuous-state systems, applying some form of binarization. The effect of binarization in the high-dimensional setup is a subject of active research: as an example, it was recently found that the angle between a random high-dimensional vector and its binarization is much smaller than typical angles between random vectors~\cite{binarization}. In regard to this, in ~\cite{Acevedo2025_2} it was found that local neighborhood ranks are approximately preserved in data space after binarization of the activations of Large Language Models.
It would thus be particularly interesting to extend our method by searching for possible relationships between the dimensionality of spin configurations and the dimensionality of binarized representations of neural networks processing, or generating~\cite{Riccardo,Hibat-Allah2021}, such configurations.

From the methodological point of view, an additional direction is the generalize of the current linear model \(d_\theta(r) = \theta_0 + r\theta_1\) in the ansatz~\eqref{eq:bid_model} to a nonlinear function that can accurately fit the full empirical distribution, even in multi-modal regimes. The dimensionality would still be extracted from \(d_\theta(r)\) but its overall shape might convey more insights about the geometry of the spin space. Preliminary results, presented in \cref{subsec:beyond_linear}, show that a cubic model for $d_{\theta}(r)$ can accurately fit empirical distributions of distances in the whole sampled regime, even in the bimodal case. Nonetheless, the numerical optimization of such a model is challenging, and we leave its systematic implementation for future work.

By bridging the gap between geometry and dynamics, the BID offers a novel perspective on the structure and behavior of complex systems, and we hope our work motivates its application in physics, neuroscience, machine learning, and beyond.
\section*{Acknowledgments}
The authors thank Federica Gerace, Marcello Dalmonte, Nina Javerzat, Matteo Negri and Alessandro Laio for useful and stimulating discussions.  

\bibliographystyle{apsrev4-2}
\bibliography{ref} 

\clearpage
\appendix
\renewcommand{\thesubsection}{\thesection.\arabic{subsection}}
\renewcommand{\thefigure}{\thesection\arabic{figure}}
\renewcommand{\thetable}{\thesection\arabic{table}}
\renewcommand{\theequation}{\thesection\arabic{equation}}

\setcounter{figure}{0}
\setcounter{table}{0}
\setcounter{subsection}{0}
\setcounter{equation}{0}

\section{Hopfield Model} \label{sec:Appendix-Hop}

\subsection{Averages}\label{subsec:averages}

In spin glasses such as the Hopfield model there are several ways in which one can define average values. To make explicit this distinction, consider a system of $N$ neurons with $p=\alpha N$ patterns $\bm \xi = \{\bm\xi ^\nu\}_{\nu=1}^p$ at temperature $T$. With a fixed realization of $\bm \xi$ we can define the \emph{quenched spin glass order parameter} as:
\begin{equation}
    q^N_{\bm \xi} = \dfrac{1}{N} \sum_{i=1}^N \langle s_i\rangle^2_{\bm \xi}
\end{equation}
where $\langle \cdot \rangle_{\bm \xi}$ is the thermal average at fixed $\xi$. 

In addition to that we can include the \emph{disorder average} as: 
\begin{equation}
    \overline{q^N} = \mathbb{E}_{\bm \xi}\left[ q^N_{\bm \xi}\right]
\end{equation}

If the system is self-averaging, in the thermodynamic limit it holds:
\begin{equation}
    q^N_{\bm \xi} \xrightarrow[N\rightarrow\infty]{P} q = \lim_{N\rightarrow\infty}\overline{q^N}
\end{equation}

Therefore $q$ is the actual spin glass order parameter in the thermodynamic limit. All this averages in practice will be estimates with a dataset $\bm S = \{\bm s^{(t)}\}_{t\leq N_S}$ collected by running \cref{alg:gibbs}, therefore we have the estimates $\widehat{\cdots}$ as: $\widehat{q_{\bm \xi}^N} \approx q_{\bm \xi}^N$ and $\widehat{\overline{q^N}} \approx \overline{q^N}$ (as in \cref{eq:order-params-estimates}). \\

One important result that we found empirically is that the estimate $\widehat{q_{\bm \xi}^N}$ does not work well in the spin glass phase for temperatures close to $T_g$ when $N$ is small. This can be observed in \cref{fig:mapping-ord-par(a)} and in the explanation of \cref{fig:scaling-N(b)}. The main observation is that the empirical value of $q$ drops to zero systematically before the critical line $T_g$, we explained this in relation to the $\mathbb{Z}_2$ symmetry of the system that averages out the value of $q$ for finite $N$. \\

Expanding the empirical average of both $q$ and the normalized overlap variable $x$ we can connect it with the result from \cref{fig:scaling-N(b)} since $ \widehat{q^N_{\bm\xi}}$ and $\mathbb{E}_{emp}[x]$ are exactly the same quantity up to a negligible $O(1/N_S)$ correction. 

\[ \widehat{q^N_{\bm\xi}} = \dfrac{1}{N} \sum_{i=1}^N \left(\dfrac{1}{N_S}\sum_{t=1}^{N_S} s_i^{(t)}\right)^2 = \dfrac{1}{N_S^2} \sum_{t,t'=1}^{N_S} \dfrac{\bm s^{(t)}\cdot \bm s^{(t')}}{N}\]

\[\mathbb{E}_{emp}[x] = \dfrac{1}{N_S(N_S-1)} \sum_{t\neq t'}^{N_S}\dfrac{\bm s^{(t)}\cdot \bm s^{(t')}}{N} \approx \widehat{q^N_{\bm\xi}}\]

The important result to notice is that the average of the overlap under the model \cref{eq:bid_model} approximates better the  actual spin order parameter $q$ and it does not suffer from the strong finite size effects.

\begin{equation}
    \mathbb{E}_{\theta}[x] \approx q
\end{equation}

\subsection{Sampling}\label{subsec:sampling}

Following the asynchronous dynamics of the Hopfield network is equivalent to sampling the Boltzmann distribution of the model using \emph{Gibbs sampling}. The sampling algorithm updates spins \textbf{one at a time} using their conditional probability (which is easily computable from \cref{eq:boltzmann-measure}):
\begin{equation}
    P(s_i = +1 \mid \bm{s}_{\setminus i}) = \frac{1}{1 + \exp(-2\beta h_i)},
    \label{eq:cond}
\end{equation}
where the local field at site \(i\) is given by
\begin{equation}
    h_i = \sum_{j\neq i} J_{ij} s_j.
    \label{eq:loc_field}
\end{equation}

To run the algorithm we started from a configuration $\bm s^{(0)}$ in which a fraction $m_0$ of spines are identical to those of $\bm\xi^0$ and the rest of them is selected randomly.\\

\begin{algorithm}
        \caption{Gibbs Sampling}
        \label{alg:gibbs}
        \KwIn{Initial Configuration: $\bm s^{(0)}\in \{-1,+1\}^N$}
        \For{$t = 1$ \KwTo $N_s$}{
            \For{$i=1$ \KwTo $N$}{
                Define $\bm{s}_{\setminus i}^{(t)} =\left(\{s_j^{(t)}\}_{j<i} \;\;, \;\;\{s_j^{(t-1)}\}_{j>i} \right)$
                
                Compute $h_i^{(t)}$ from  \cref{eq:loc_field} using $\bm{s}_{\setminus i}^{(t)}$
                
                Update $s_i^{(t)}$ based in \cref{eq:cond} using $h_i^{(t)}$
                }
            Transition $\bm s^{(t-1)} \rightarrow\bm s^{(t)}$
            }
        \KwOut{Dataset $\bm S = \{\bm s^{(t)}\}_{t\leq N_S}$}
\end{algorithm}

The initialization with different values of $m_0$ is crucial to be able to uncover the retrieval-metastable phase, since in this region the magnetized state is not a global minimum of the free energy and the dynamics converges to one of the patterns only if we are initially in its basin of attraction, otherwise the dynamics ends up in a glassy state.\\

\subsection{Order Parameters}\label{subsec:order_parameters}

\begin{figure}[H]
    \begin{subfigure}[t]{1\linewidth} 
        \caption{ \raggedright }
        \includegraphics[width=0.9\linewidth]{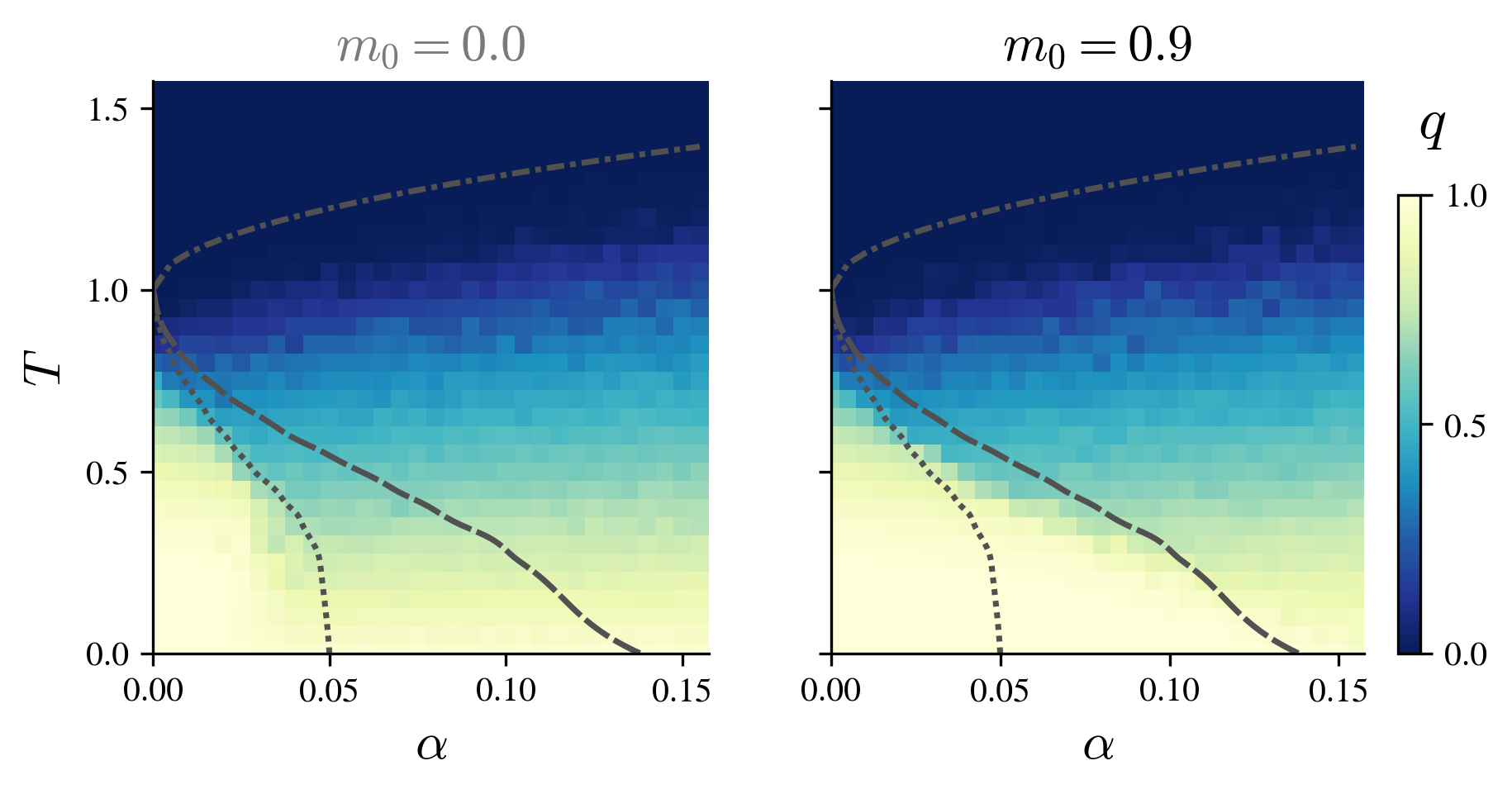}
        \label{fig:mapping-ord-par(a)}
    \end{subfigure}

    \begin{subfigure}[t]{1\linewidth}
        \caption{\raggedright}
        \includegraphics[width=0.9\linewidth]{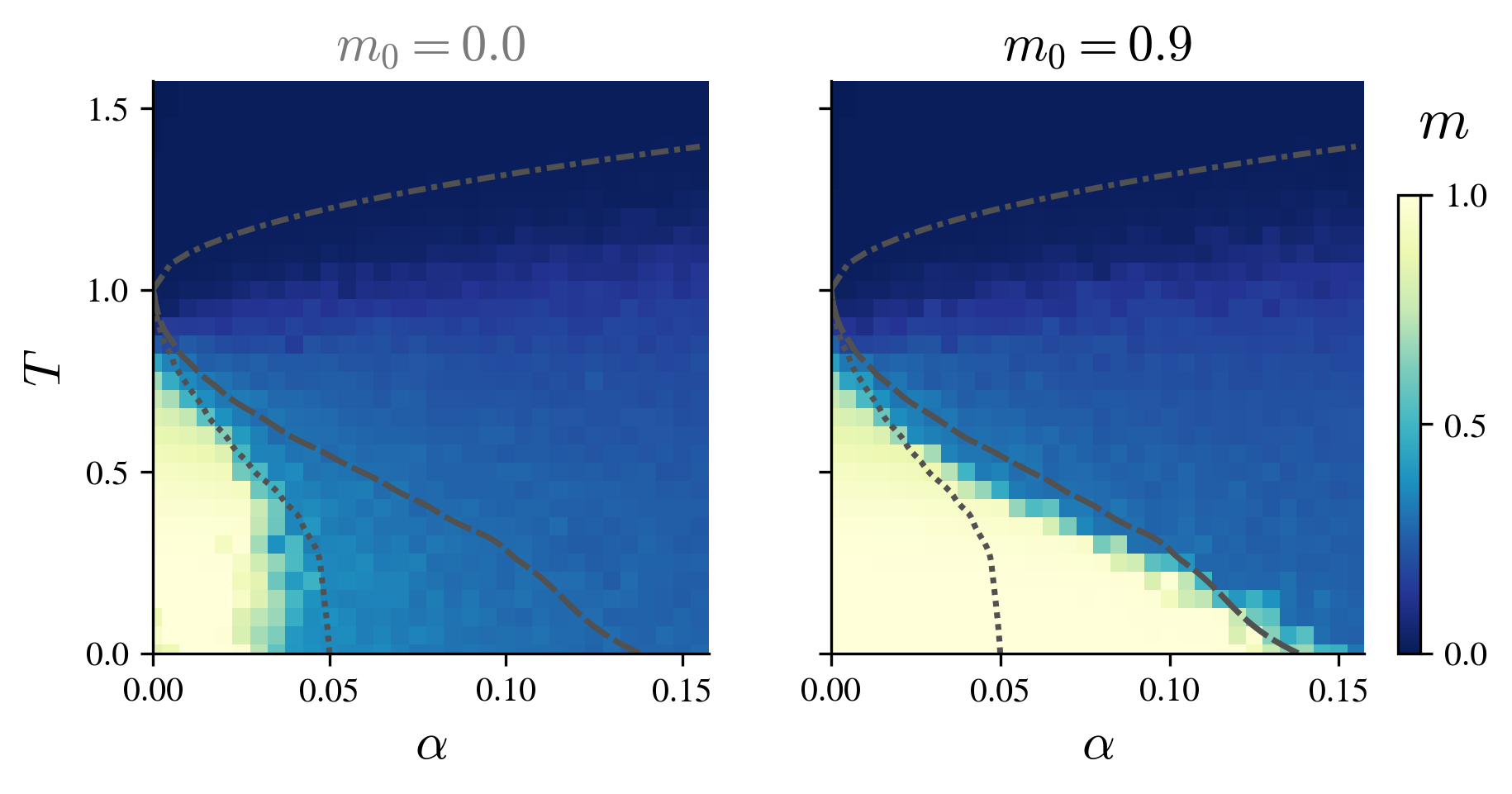}
        \label{fig:mapping-ord-par(b)}
    \end{subfigure}

\caption{Spin order parameters across space \((\alpha,T)\) estimated using \cref{eq:order-params-estimates}. (a) $\hat q$ and (b) $\hat m$. Simulations were performed with $N=1024$, $N_S=2000$ and average values over $10$ realizations.}
\label{fig:mapping-ord-par}
\end{figure}

Along with the map of the $BID$ in \cref{fig:BID-ord-param(a)} we also estimated the Hopfield order parameters across the $(\alpha,T)$ space in \cref{fig:mapping-ord-par}. We can qualitatively distinguish some regimes in the plot which correspond to the different phase transitions. The metastable region is clear looking at $m$ when we pass from $m_0=0$ to $m_0 = 0.9$. In this retrieval region for both $m_0$ the magnetization $m$ is close to $1$ and it drops quickly when we enter in the spin glass phase.\\

The theory of the Hopfield network in the thermodynamic limit indicates that in the spin glass and paramagnetic phases the overlap \(m\) should be $0$, but for finite size simulations that is not the case. In \cref{fig:mapping-ord-par(b)} we see that across the spin glass phase $m$ is slightly greater than zero and it only vanishes once $q$ also goes to zero. This finite size effect was more evident in \cref{fig:scaling-N(a)} where the value of $m$ slowly converges to $0$ in the spin glass phase when we increase $N$. The fact that the value of $q$ drops to zero before the transition temperature $T_g$ indicates that the transitions between $\mathbb{Z}_2$ symmetrical states is taking place which cancels the empirical average in \cref{eq:order-params-estimates} for $\hat{q}$.

\subsection{Replica Solution}\label{subsec:replica_solution}

The transition temperatures from \cref{fig:BID-ord-param,fig:mapping-ord-par} were found solving numerically the saddle point equations for the Replica Symmetric free energy of the Hopfield Network \cite{Nishimori_2001}.

\begin{equation}
\begin{split}
    m &= \int Dz \; \tanh \beta (\sqrt{
    \alpha r} z +m)\\
    q &= \int Dz \; \tanh^2 \beta (\sqrt{
    \alpha r} z +m)\\
    r&=\dfrac{q}{(1-\beta+\beta q)^2}
\end{split}
\label{eq:RS_fixed_point_equations}
\end{equation}

The temperature for the paramagnetic phase transition can be obtained analytically as \(T_g = 1+\sqrt{\alpha}\) by expanding \cref{eq:RS_fixed_point_equations} with small \(q\) and \(m=0\). Above \(T_M\) only the Spin Glass (SG) solution (with \(m=0\)) exists. Below \(T_M\) there are two solutions: retrieval and spin glass, and the critical temperature \(T_c\) is obtained when the free energy evaluated at both solutions is the same.

\section{Binary Intrinsic Dimension} \label{sec:Appendix-BID}

\subsection{Gaussian Approximation} \label{subsec:Gaussian_apx}

The model for the distribution of Hamming distances according to \cite{Acevedo_2025} is: 
\begin{equation}
    P(r|\theta) = \dfrac{1} {Z} \dfrac{1}{2^{d_{\theta}(r)}} \binom{d_\theta(r)}{r} 
    \label{eq:apendix_model_P_theta}
\end{equation}
with 
\[
    \quad d_\theta(r) = \theta_0 + \theta_1 r \quad \text{and} \quad Z = \displaystyle\sum_{r\in\mathcal{R}} \dfrac{1}{2^{d_{\theta}(r)}} \binom{d_\theta(r)}{r}
\]

The full range of possible distance values is $\mathcal{R} = \{0,1,\cdots,N\}$ where $N$ is the size of the system, but in practice \(\mathcal{R}\) is the range of values observed in the empirical distribution, or a subset of it to consider only the portion of the empirical distribution that we are interested in. 

To perform the approximation of the model, we need to write a Taylor expansion of $\log P$ around its maximum. For this, we consider it as a function of \(x = 1-\frac{2}N{r}\) and we assume it is a continuous variable for large \(N\). To this end we first extend the definition of the binomial coefficient using the gamma function:

\[
    \displaystyle\binom{d_{\theta}(r)}{r} \equiv \dfrac{\Gamma(d_{\theta}(r)+1)}{\Gamma(d_{\theta}(r)-r+1)\;\; \Gamma(r+1)}.
\]

Using the property that \(\frac{d\log \Gamma(z+1)}{dz}\approx\log z\) we can write:

\begin{align*}
    \dfrac{d}{dr} \log P(r |\theta) = -\theta_1 \log 2 + \theta_1 \log d_\theta(r) \\
    - (\theta_1-1)\log(d_\theta(r)-r) - \log r
\end{align*}

To find the maximum $r_0$ we set $ \dfrac{d}{dr} \log P(r=r_0 |\theta) = 0$ and we will see that the solution is: 
\[
    r_0 = \dfrac{\theta_0}{2-\theta_1}
\]

for this solution to be inside of the range $\mathcal{R}$ we require $r_0 < N$ and this will give the condition: 
\begin{equation}
    \theta_0 < N(2-\theta_1).
    \label{eq:constrain}
\end{equation}

With the expression above we can take higher order derivatives: 

\begin{align*}
    \dfrac{d^2}{dr^2} \log P(r |\theta) &= \dfrac{\theta_1^2}{d_\theta(r)} - \dfrac{(\theta_1-1)^2}{d_\theta(r)-r}-\dfrac{1}{r} \\
    \Rightarrow \quad \dfrac{d^2}{dr^2} \log P(r&=r_0 |\theta) = -\dfrac{(2-\theta_1)^3}{2\theta_0}
\end{align*}

This expansion allows us to write the distribution as: 

\begin{equation}
    P_{gauss}(r|\theta) = \dfrac{1}{\tilde Z} \exp\left(-\dfrac{(r-r_0)^2}{2\sigma_r^2} \right)
    \label{eq:apx_P_gauss}
\end{equation}

where
\begin{equation}
    r_0 = \dfrac{\theta_0}{2-\theta_1}  \; \; \; \text{and}\; \; \;   \sigma_r^2 = \dfrac{2 \theta_0}{(2-\theta_1)^3}.
\end{equation}

and the reverse transformation of parameters is: 
\begin{equation}
    \theta_0 = \sqrt{\dfrac{2\;r_0^3}{\sigma_r^2}}\qquad \text{and}\qquad \theta_1 = 2-\sqrt{\dfrac{2r_0}{\sigma_r^2}}.
    \label{eq:thetas_gauss_apx_init}
\end{equation}

The normalization should be computed numerically, although in regions of the parameter space where the constraint \cref{eq:constrain} holds we actually have: 
\[
    \tilde Z = \displaystyle\sum_{r\in \mathcal{R}} \exp\left(-\dfrac{(r-r_0)^2}{2\sigma_r^2} \right) \approx \sqrt{2 \pi \sigma_r^2}
\]

\begin{figure}[H]
    \centering
    \includegraphics[width=0.8\linewidth]{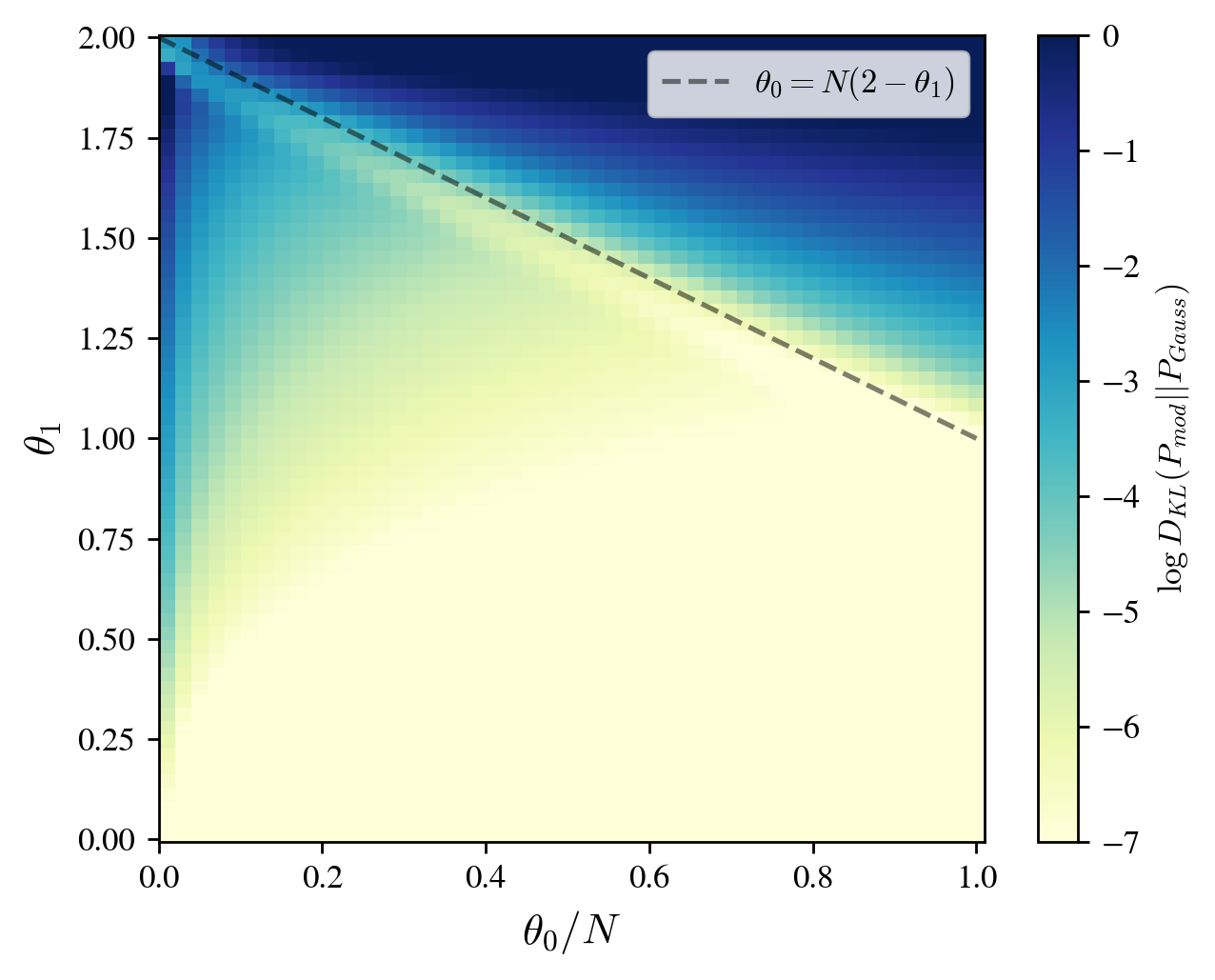}
    \caption{Comparison between models in terms of $\log D_{KL}$. $N=1000$. We evaluated the model \cref{eq:bid_model} and the gaussian approximation \cref{eq:apx_P_gauss} with the same parameters \(\theta_0,\theta_1\) and computed the divergence between these distributions.}
    
    \label{fig:compare_gauss_model}
\end{figure}

In \cref{fig:compare_gauss_model} we can see that the Gaussian model is a good approximation of \cref{eq:bid_model} for a wide range of parameters. The approximation gets significantly worse above the black dashed line because, for that region, the peak of the distribution goes outside of the range of possible values for \(r\).

\subsection{Gradient}\label{subsec:gradient}

The Kullback-Leibler divergence between the empirical distribution and the model is:

\begin{equation}
    D_{KL}(\theta)  = \displaystyle\sum_{r\in\mathcal{R}} P_{emp}(r) \log \left(\dfrac{P_{emp}(r)}{P(r|\theta)}\right).
\end{equation}

To find the optimal parameters via gradient descent, we need to compute the derivatives of $D_{KL}(\theta)$ and follow the standard update rule with an appropriate learning rate $\eta$: 
\begin{equation}
    \theta \leftarrow \theta - \eta \cdot \nabla D_{KL}(\theta).
    \label{eq:A.2_gradient_descent}
\end{equation}
To compute the gradient it is useful to define some auxiliary functions first. From the binomial coefficient, let  us define:
\begin{equation*}
\begin{split}
    f_\theta(r) \equiv \dfrac{\Gamma(d_{\theta}(r)+1)}{\Gamma(d_{\theta}(r)-r+1)} & = \displaystyle\prod_{l=0}^{r-1}(d_{\theta}(r)-l) \;\;\;\text{for}\;\;\; r>0\\
    \text{and}\;\;\; f_\theta(r) &= 1  \;\;\; \text{for}\;\;\; r=0. 
\end{split}
\end{equation*}
With this definition we have:
\[
    \log f_\theta(r) =  \displaystyle\sum_{l=0}^{r-1} \log \left(d_{\theta}(r)-l) \right) 
\]
and therefore \( \nabla \log f_\theta(r) = \frac{\nabla f_\theta(r)}{f_{\theta}(r)} = S^1_{\theta}(r) \; \nabla d_{\theta}(r) \), where the new function $S^\lambda_\theta(r)$ is defined as: 
\begin{equation}
    S^\lambda_\theta(r) \equiv  \displaystyle\sum_{l=0}^{r-1} \dfrac{1}{(d_{\theta}(r)-l)^\lambda}
    \label{eq:A.2_S_alpha}
\end{equation}
With the new definitions we can write:
\[
\nabla\log P(r|\theta) =  \left( S^1_{\theta}(r)-\log 2 \right)\nabla d_{\theta}(r) - \nabla \log Z(\theta)
\]
where:
\begin{equation}
        \nabla \log Z(\theta) = \displaystyle\sum_{r\in\mathcal{R}} P(r|\theta) \left( S^1_{\theta}(r)-\log 2 \right)\nabla d_{\theta}(r)
\end{equation}

and finally compute the gradient:

\begin{equation}
    \begin{split}
        \nabla & D_{KL}(\theta) = -\displaystyle\sum_{r\in\mathcal{R}} P_{emp}(r) \nabla \log P(r|\theta) \\
     = \displaystyle\sum_{r\in\mathcal{R}}&  \left[P_{emp}(r)  -P(r|\theta)\right] \left[\log 2 - S^1_{\theta}(r)\right]\nabla d_{\theta}(r)   
    \end{split}
     \label{eq:A.2_gradient}
\end{equation}

After computing the gradient \(\nabla D_{KL}(\theta)\), the selection of the learning rate is crucial to achieve fast convergence in \cref{eq:A.2_gradient_descent}. The advantage of this model is that it allows us to use a second order method for the optimization since we can explicitly compute the second derivatives of the \(D_{KL}(\theta)\). \\
For this reason we can choose the learning rate as the inverse Hessian matrix of the Kullback-Leibler divergence.
\begin{equation}
    \eta  \equiv \nabla^2 D_{KL}(\theta)^{-1}
\end{equation}

The final result can be written only in terms of the auxiliary functions \(S^\lambda_\theta(r)\), and the gradient and hessian of the dimensionality model (\(\nabla d_\theta(r),\;\nabla^2 d_\theta(r)\)).

\begin{widetext}

\begin{equation}
\begin{split}
\nabla^2 D_{KL}(\theta) = \displaystyle\sum_{r\in\mathcal{R}} \Bigl\{\;P(r|\theta)\left( \log 2 - S^1_\theta(r) \right)^2
+\left( P_{emp}(r) - P(r|\theta)\right)S^2_\theta(r)\;\Bigr\} \nabla^T d_\theta(r)\nabla d_\theta(r)   \\
+\displaystyle\sum_{r\in\mathcal{R}}  \left( P_{emp}(r) - P(r|\theta)\right)\left( \log 2 - S^1_\theta(r) \right) \nabla^2 d_\theta(r) \;-\;\nabla^T \log Z(\theta)\nabla \log Z(\theta)\\ 
\end{split}
 \label{eq:A.2_hessian}
\end{equation}

\end{widetext}

With \cref{eq:A.2_gradient,eq:A.2_hessian} we can perform an extremely fast and accurate gradient descent algorithm to find the optimal set of parameters $\hat{\theta}$. For this optimization to work is crucial the correct initialization of the parameters and for this reason the Gaussian approximation is also practically useful. To initialize the parameters, we computed the mean and the variance of the empirical distribution and use \cref{eq:thetas_gauss_apx_init} to obtain an educated guess for 
$\theta$. This drastically improves the descent algorithm over random initializations which can lead to unstable or slow descent dynamics.\\
It is important to emphasize also that \cref{eq:A.2_gradient,eq:A.2_hessian} are general expressions written in terms of the gradient and hessian of the dimensionality model (\(\nabla d_\theta(r),\;\nabla^2 d_\theta(r)\)) without explicit reference to the linear ansatz used by Acevedo et al. \cite{Acevedo_2025}. This opens the possibility for generalizations of the current model $d_\theta(r) = \theta_0 + r\theta_1$ that can be implemented easily within the same numerical setup.

\subsection{Selection of the range $\mathcal{R}$.}\label{subsec:range_R}

To perform the optimization, we fit the distribution of Hamming distances in a restricted range \( \mathcal{R} = \{r_{min},r_{max}\} \) that we defined based on the percentiles of the distribution (\( \alpha_{min},\alpha_{max} \)) defined as:
\[\alpha_{min/max} = \sum_{r=0}^{r_{min/max}} P_{emp}(r)\]

\begin{figure}[H]
    \centering
    \includegraphics[width=1\linewidth]{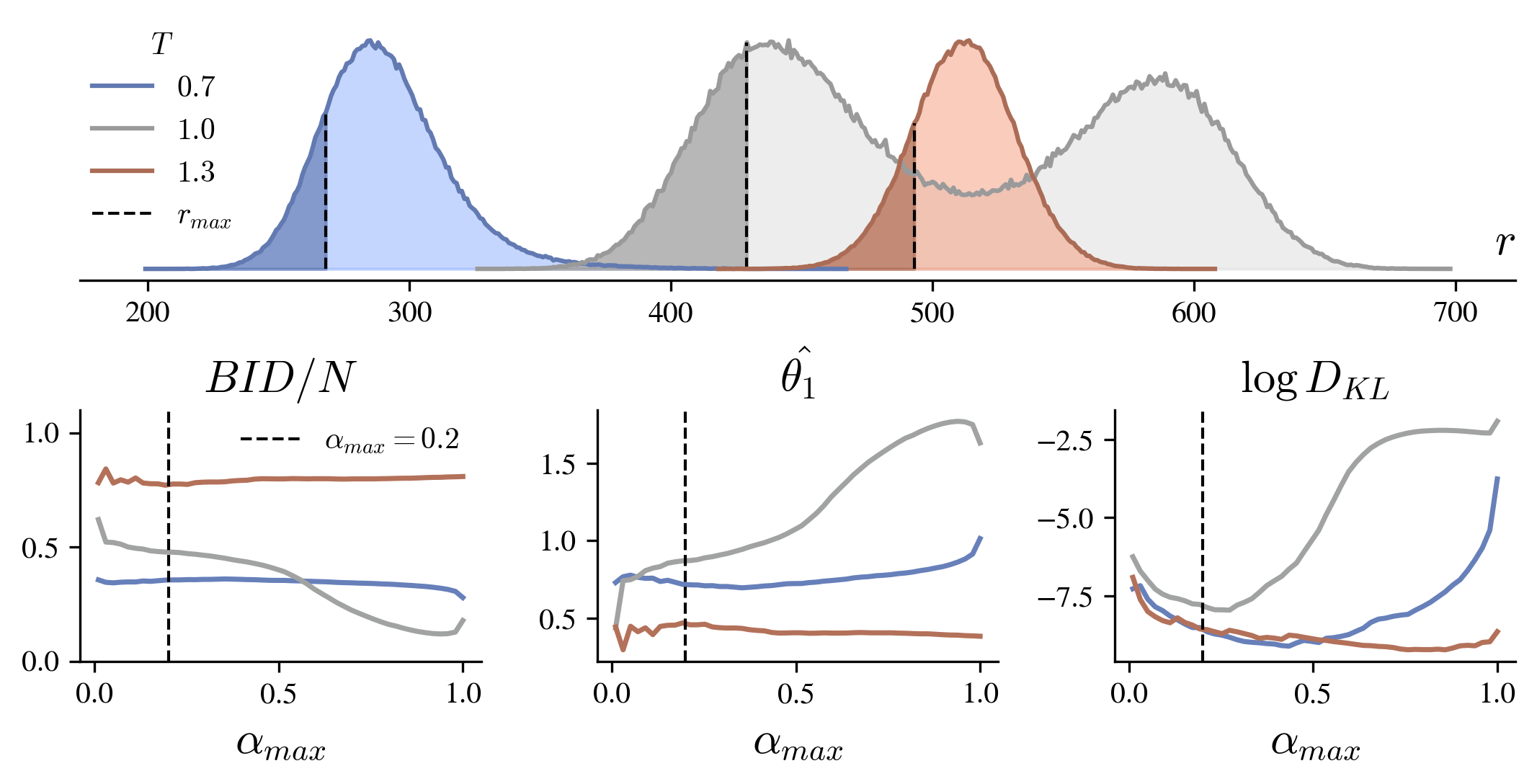}
    \caption{Top: empirical histogram of distances at different temperatures. The darker region indicates the range \(\mathcal{R}\) obtained with \((\alpha_{min},\alpha_{max}) = (0.0,0.2)\). Bottom: parameters obtained after fitting the model to the empirical distribution which is truncated according to the value of \(\alpha_{max}\). The value of \(\log D_{KL}\) is also shown as a function of \(\alpha_{max}\). Simulations were performed with $N=1024$, $N_S=1000$ and $\alpha=p/N=0.04$, \(\alpha_{min} = 0.0\) .}
    \label{fig:alpha_max_paramns}
\end{figure}

In \cref{fig:alpha_max_paramns} we show that the estimated parameters \(\hat\theta\) are very robust under the selection of the interval when the distribution is unimodal. When the distribution is bimodal and we increase \(\alpha_{max}\) further than \(\approx 0.5\) the value of \(\log D_{KL}\) increases rapidly signaling the poor fit of the distribution, and also the estimated parameters change. 
In practice we used a value \(\alpha_{max} = 0.2\) in the simulations because it performed well in both uni-modal and bimodal cases. As mentioned in \cref{sec:BID-and-q}, we focused on the mode of small distances when the distribution is bimodal because it contains intra-group distances which allows us to work as if there was only a single cluster of states in the spin space with a common geometry.

\begin{figure}[H]
    \centering
        \includegraphics[width=1\linewidth]{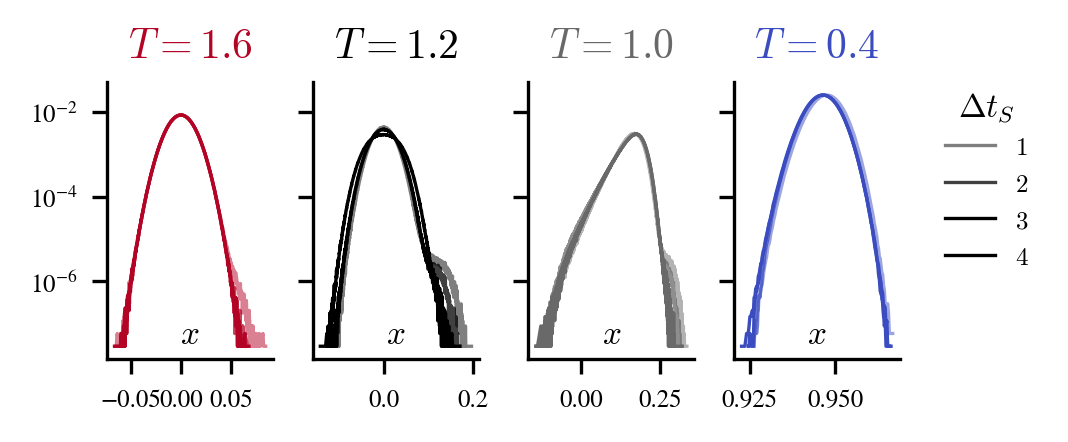}
    \includegraphics[width=1\linewidth]{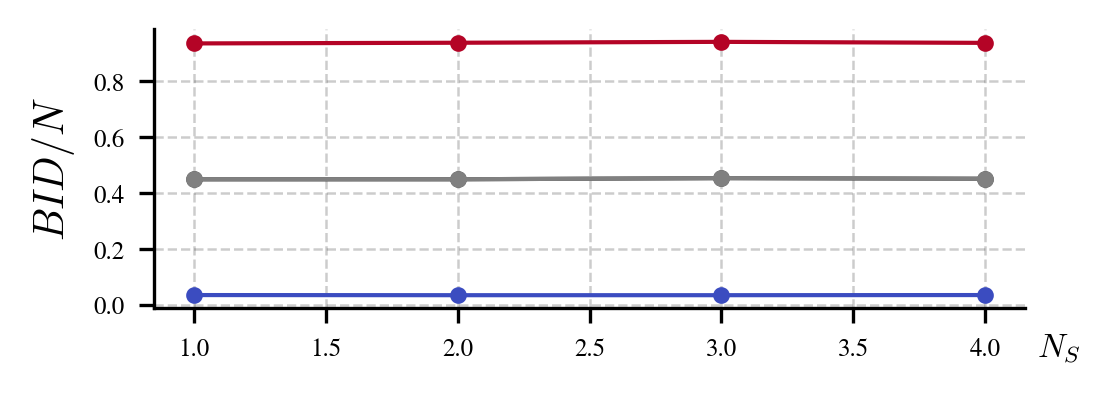}
    \caption{\raggedright The BID is robust against time correlations effects. Top: histograms at different temperatures with various sampling time intervals \(\Delta t\). When \(\Delta t\) is small there is a \emph{shoulder} on the distribution that appears which is related to correlations between consecutive samples. A proof of this is the fact that increasing the sampling time interval eliminates the \emph{shoulder}. Bottom: the estimated value of the BID is virtually blind to the time correlations since during the fitting procedure we select the range to cut out the noise at small distances. Simulations with \(N = 8192\) and \(\alpha = 0.04\).}
    \label{fig:time_correlations}
\end{figure}
The selection of the range \(\mathcal{R}\) has another advantage which is avoiding time correlations between data points. When running the \cref{alg:gibbs} to sample the spin states, consecutive samples might have strong time correlations that are inherent to the dynamics of the system. In principle the time correlations can be avoided if we increase the sampling time interval, but in practice this is computationally prohibitive for large systems. In \cref{fig:time_correlations} we can see how the BID gives a robust estimation even in the presence of these correlations.

\subsection{Scaling Laws} \label{subsec:scaling_laws}
\begin{figure}[H]
    \begin{subfigure}[ht]{0.5\linewidth}
    
        \centering
        \includegraphics[width=1.\linewidth]{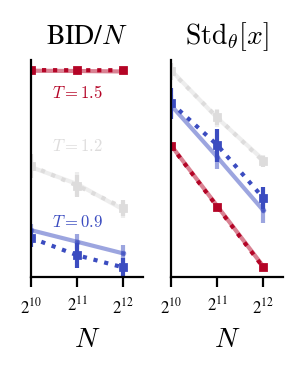}
    \end{subfigure}
    \hfill
    \begin{subfigure}[ht]{0.48\linewidth}
   
    \centering Single chain
        \begin{tabular}{|c|c|c|}
                \hline
                 $T$& $\gamma_{BID/N}$ & $\gamma_{\text{Std}_\theta[x]}$ \\\hline
                $1.5$ & $0.07 \pm 0.04$ & $0.43\pm0.07$\\\hline
                 $1.2$ & $0.14\pm0.03$ & $0.37\pm0.03$\\\hline
                 $0.9$ & $0.0\pm0.002$ & $0.5\pm0.002$\\\hline
            \end{tabular}\\
            
     \centering Multi chain
        \begin{tabular}{|c|c|c|}
                \hline
                 $T$& $\gamma_{BID/N}$ & $\gamma_{\text{Std}_\theta[x]}$ \\\hline
                $1.5$ & $0.1 \pm 0.04$ & $0.4\pm0.08$\\\hline
                 $1.2$ & $0.14\pm0.04$ & $0.37\pm0.04$\\\hline
                 $0.9$ & $0.0\pm0.004$ & $0.5\pm0.003$\\\hline
            \end{tabular}
    \end{subfigure}
    \caption{Comparison of the scaling exponents obtained with different sampling schemes. Simulation with \(\alpha=0.04\).}
    \label{fig:scale_bid_sigma}
\end{figure}

In~\cref{sec:scaling_exp} we observed that in the whole spin glass phase the BID scales sublinearly with system size. To be sure that this is not a trivial effect of the unavoidable time autocorrelation present in the system, we ran two types of dynamics: 
In single chain (solid lines in~\cref{fig:scale_bid_sigma}) the samples are gathered running one dynamics for a long time, starting from a single initial condition. In multi chain (dotted lines) the samples are gathered all from independent chains with random initial conditions. The table in~\cref{fig:scale_bid_sigma} shows the corresponding scaling exponents, which are in very good agreement.

\subsection{Beyond Linear Model}\label{subsec:beyond_linear}

In the main text, we argue that the linear model \(d_\theta(r) = \theta_0 + r\theta_1\) proposed in \cite{Acevedo_2025} allows us to fit very well the empirical distribution of the data. Nonetheless, when the empirical distribution is very asymmetric or bimodal, such linear model can only fit the it with~\cref{eq:bid_model} on a restricted range of distances (see \cref{subsec:range_R}). 
Then, we can imagine that if we use a more flexible model for the dimensionality \(d_\theta(r)\) we will be able to fit empirical distributions with arbitrarily complex shapes. The simplest generalization of the linear model is the following polynomial dimensionality of degree \(k\):

\begin{equation}
 d^{(k)}_{\theta}(r) = N  \sum_{n=0}^k \theta_n \left(\dfrac{r}{N}\right)^n \;\text{ where } \; \theta = (\theta_n)_{n\leq k}.
 \label{eq:poly_model}
\end{equation}

\cref{eq:poly_model} reduces to the linear case when \(k=1\) (up to a rescaling of the zero order coefficient) and it is very easy to test since the optimization algorithm (\cref{subsec:gradient}) is written for a general dimensionality model.\\

\begin{figure}[H]
    \begin{subfigure}[ht]{0.54\linewidth}
        \includegraphics[width = 1.0\linewidth]{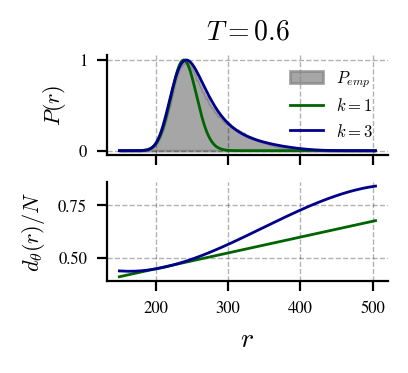}
    \end{subfigure}
    \begin{subfigure}[ht]{0.44\linewidth}
        \includegraphics[width = 1.0\linewidth]{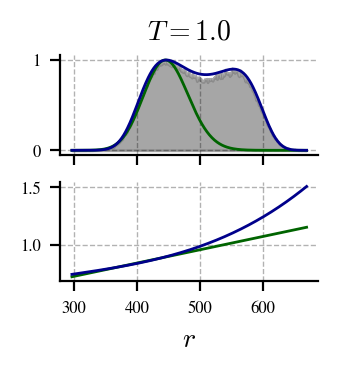}
    \end{subfigure}
    \caption{ Comparison between linear and polynomial models. For both temperatures, the empirical distribution is either very non symmetrical or multi-modal which causes the linear model to fail. The polynomial model is able to fit all the distributions almost perfectly. Simulation with \(N=1000\), \(\alpha=0.04\).}
    \label{fig:poli_model_hist}
\end{figure}

In \cref{fig:poli_model_hist} we contrast the results obtained with a linear model (\(k=1\)) versus the ones obtained with a polynomial of degree \(k=3\). The first clear observation is that the higher degree model is able to fit \(P_{emp}\) almost perfectly in both temperatures whereas the linear case only works well at small distances. Interestingly, the shape of the dimensionality measure \(d_\theta(r)\) in both models is very similar for the smallest sampled values of \(r\) where the measured histogram has a positive slope, but the polynomial model differs for larger distances. 

The results of the fitting algorithm are:
\begin{align*}
    T&=0.6\\
    \theta_{k=1} &= (0.298,0.753)\\
    \theta_{k=3} &= (0.753,-4.3,17.16,-16.44)\\
        \text{}\\
    T&=1.0\\
    \theta_{k=1} &= (0.374,1.16)\\
    \theta_{k=3} &= (0.305,2.86,-7.18,8.35)\\
\end{align*}

In the linear case (\(k=1\)), the first parameter corresponds to the value of \(BID/N\) which does not match the parameter \(\theta_0\) of the polynomial case. This shows that with this vanilla approach, the two models have different limiting values for $r\rightarrow0$, and we leave for future work how to reconcile them, or how and why one should choose one over the other.

The extension to non-linear models opens the possibility of studying the full curve \(d_\theta(r)\) itself to extract information about the data and eliminates the necessity of pre-processing the histogram before performing the fit, which results in a more robust methodological tool for data analysis. We note, however, that in the right panel of~\cref{fig:poli_model_hist} both models for $d_{\theta}(r)$ give values larger than the number of spins in the system $(N)$ for scales r belonging to the right hand side probability peak. This could mean that there are other solutions which are instead well behaved, having values of $d_{\theta}(r)$ which could be interpreted as a scale dependent intrinsic dimension, necessarily smaller than the number of degrees of freedom. 
\end{document}